\pdfoutput=1
\RequirePackage{fix-cm}
\documentclass[twocolumn]{svjour3}          % twocolumn

\smartqed  % flush right qed marks, e.g. at end of proof
\usepackage{graphicx}
\usepackage{natbib}
\usepackage{multirow}
\usepackage{amsmath, amssymb}
\usepackage{color}

\journalname{Submitted}
\begin{document}

\title{Reverse Engineering Gene Regulatory Networks Using Approximate Bayesian Computation%\thanks{Grants or other notes
}

\titlerunning{Reverse Engineering Gene Networks Using ABC}        % if too long for running head

\author{Andrea Rau         \and
        Florence Jaffr{\'e}zic 	\and
        Jean-Louis Foulley		\and
        R. W. Doerge
}


\institute{A. Rau \and R. W. Doerge \at
              Department of Statistics, Purdue University, West Lafayette, IN, USA	
           \and
           A.Rau \and F. Jaffr{\'e}zic and J.-L. Foulley \at
			INRA, UMR 1313 GABI, Jouy-en-Josas, France
		   \and
		   R. W. Doerge \at
		      Department of Agronomy, Purdue University, West Lafayette, IN, USA\\	
		      Tel.: +001-765-494-3141\\
              Fax: +001-765-494-0558\\
           \email{doerge@stat.purdue.edu}           	
}

\date{September 7, 2011}

\maketitle

\begin{abstract}
Gene regulatory networks are collections of genes that interact with one other and with other substances in the cell. By measuring gene expression over time using high-throughput technologies, it may be possible to reverse engineer, or infer, the structure of the gene network involved in a particular cellular process. These gene expression data typically have a high dimensionality and a limited number of biological replicates and time points. Due to these issues and the complexity of biological systems, the problem of reverse engineering networks from gene expression data demands a specialized suite of statistical tools and methodologies. We propose a non-standard adaptation of a simulation-based approach known as Approximate Bayesian Computing based on Markov chain Monte Carlo sampling. This approach is particularly well suited for the inference of gene regulatory networks from longitudinal data. The performance of this approach is investigated via simulations and using longitudinal expression data from a genetic repair system in {\it Escherichia coli}.
\keywords{Approximate Bayesian computation \and Gene regulatory networks \and Longitudinal gene expression \and Markov chain Monte Carlo}
\end{abstract}

\section{Introduction}
\label{s:intro}
\begin{figure}[t!]
	\label{geneNetConcept}
  \centering
  \includegraphics[width=.5\textwidth, clip = true, trim = .1cm .1cm 5cm 19cm]{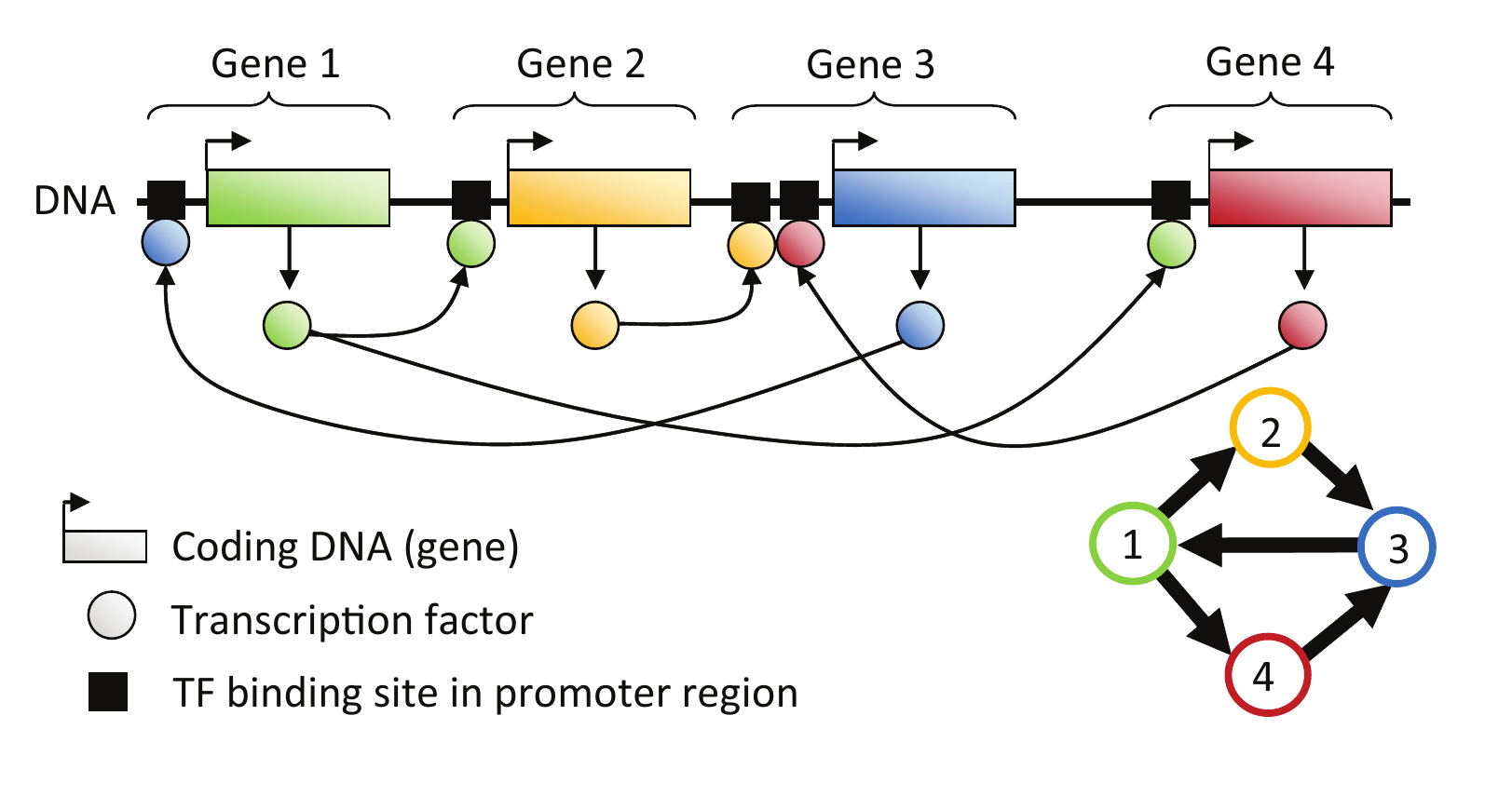}
  \caption{A simple gene regulatory network made up of four genes. Each gene is transcribed and translated into a transcription factor (TF) protein, which in turn regulates the expression of other genes in the network by binding to their respective promoter regions \citep{Schlitt2007}. The gene regulatory network may be represented using the graph in lower right corner, made up of four nodes (genes) and five edges (interactions among the genes). Image taken from \citet{Rauthesis2010}}
  \label{fi:GeneNet}
\end{figure}

%	- Gene expression (high-throughput technologies, e.g., microarrays), time-course data
The development of high-throughput technologies, such as microarrays and next-generation sequencing, has enabled large-scale studies to simultaneously assay the expression levels of thousands of genes over time. However, in spite of the abundance of data obtained from these technologies, it can be very difficult to unravel the patterns of expression among groups of genes, often referred to as gene regulatory networks \citep[GRN;][]{Friedman2004, Wilkinson2009}. Within GRN, genes interact with one another indirectly through proteins known as transcription factors (TF), which control the transfer of information during transcription by activating or repressing a ribonucleic acid (RNA) polymerase, and thus affect the level of gene expression (Figure \ref{geneNetConcept}). A GRN can thus be described as the interactions that occur (indirectly through messenger RNA and TF) within a collection of interconnected genes. 

%	- Complications of inferring networks (P >> N), why Bayesian methods and DBN (cite SSM) are well-suited
In longitudinal studies of gene expression, the number of samples (i.e., biological replicates or time points) collected is typically far outweighed by the number of observed genes. Because of the exponentially large number of possible gene-to-gene interactions, reverse engineering GRN from longitudinal studies actually amplifies the ``large $p$ small $n$" paradigm. In addition, because gene-to-gene interactions coincide with reactions in the cellular environment, the network structure can itself be very complex. For this reason, standard statistical techniques cannot be used to infer GRN from gene expression data, and a specialized suite of statistical methodologies have been developed. Among these methods, a framework known as Dynamic Bayesian Networks (DBN) has seen wide application in the context of GRN \citep{Husmeier2003, Rangel2004, Beal2005, Rau2010}. A DBN uses time-series measurements on a set of random variables to characterize their interactions over time. To avoid an explosion in model complexity, a time-homogeneous Markov model is typically used \citep{Husmeier2005}. This restriction implies that gene-to-gene interactions are constant across time and that biological samples are taken at equidistant time points.

%	- Background of Bayesian methods and why they are appropriate
Within the framework of DBN, the Bayesian paradigm is particularly well-suited to the inference of GRN for a number of reasons. First, the number of possible network structures increases exponentially as the number of genes increases \citep{Husmeier2005}. As a large number of network structures may yield similarly high likelihoods, attempting to infer a single globally optimal structure may be meaningless. In such cases, posterior distributions of gene-to-gene interactions may better characterize a GRN. Second, by examining the shape of the posterior distributions within portions of a GRN, additional information may be gleaned about the structure and inferability of specific gene-to-gene interactions, as well as the system as a whole. Finally, a Bayesian framework allows \textit{a priori} knowledge to be encoded in the prior distribution structure. Prior knowledge may refer to certain features of the topology of a GRN (e.g., sparsity in the network structure or the maximum number of regulators per gene) and to prior biological information about well-characterized pathways from bioinformatics databases.

%	- Background of ABC methods and why they are appropriate
Unless restrictive assumptions are made about the dynamics of the system (e.g., Gaussian prior distributions for gene-to-gene interactions), the likelihood function of a GRN may be intractable or difficult to calculate. In such cases, sampling-based approximate Bayesian computation (ABC) methods can allow Bayesian inference to be adopted \citep{Pritchard1999, Beaumont2002, Marjoram2003} when simulation from the model is straightforward. The first implementation of an ABC algorithm was introduced by \citet{Pritchard1999}. In this approach, using parameter values simulated from a prior distribution, data are simulated and compared to the observed data. When the simulated and observed data are sufficiently ``close", as determined by a distance function $\rho(\cdot)$ and tolerance $\epsilon$, the parameter values are accepted \citep{Beaumont2002}. The algorithm is approximate when $\epsilon > 0$, and its output amounts to simulating from the prior when $\epsilon\rightarrow\infty$. For $0 < \epsilon < \infty$, the algorithm results in a sample of parameters from an approximate posterior distribution. 

%	- ABC variants and lit review
Because a naive application of ABC methods can be time-consuming and inefficient, a variety of extensions have been proposed in recent years. For high-dimensional data, \citet{Beaumont2002} found that using summary statistics to compare simulated and observed data, rather than the data points themselves, enables a reduction of the data without negatively impacting the approximation. Several adaptations of ABC algorithms have also been proposed based on Monte Carlo techniques. For instance, \citet{Marjoram2003} extended the ABC algorithm to work within the Markov chain Monte Carlo (MCMC) framework without the use of likelihoods. In this approach, which we refer to as ABC-MCMC, parameters are proposed from a transition distribution (e.g., a random walk) and subsequently used to simulate data. \citet{Sisson2007} used a Sequential Monte Carlo technique (SMC-ABC) to propagate a population of parameters through a sequence of intermediary distributions to obtain a sample from the approximate posterior distribution. In related work \citet{Beaumont2009} applied an adaptive sequential technique known as Population Monte Carlo (PMC) to the general ABC algorithm to improve its efficiency through iterated importance sampling. Further recent implementations of ABC algorithms can also be found in, e.g., \citet{Leuenberger2009} and \citet{Drovandi2010}.

% Say something about how results from ABC-Net can help clear up information gleaned from other methods...
In this work, we propose an extension of the ABC-MCMC algorithm to enable the inference of GRN from time-course gene expression data. Our approach enables Bayesian inference without restrictive assumptions about the distribution of gene-to-gene interactions within a network. The resulting approximate posterior distributions of interactions within the network have the added advantage of providing salient information about the inferability of the biological system as a whole. Although there have been some recent developments in network inference using ABC methods, e.g., to compare the evolution of protein-protein interaction networks \citep{Ratmann2007, Ratman2009} and to conduct model selection for systems based on ordinary differential equations \citep{Toni2009, Toni2010}, to our knowledge this is the first application of ABC methods in reverse engineering the unknown structure of a gene regulatory network from gene expression data.

%%%%%%%%%%%%%%%%%%%%%%%%%%%%%%%%%%%%%%%%%%%%%%%%%%%%%%%%%%%%%%%%%%%%%%%%%%%%%%%%%%%%%%%%%%%%%%%%%%%%%%%%%%%%%%
%%%%%%%%%%%%%%%%%%%%%%%%%%%%%%%%%%%%%%%%%%%%%%%%%%%%%%%%%%%%%%%%%%%%%%%%%%%%%%%%%%%%%%%%%%%%%%%%%%%%%%%%%%%%%%

\section{Approximate Bayesian Computation for Networks}
\label{s:ABCNet}

Let $Y$ be a set of observed gene expression data for $P$ genes at $T$ equally-spaced time points, where $\mathbf{y}_{t} = \left( y_{1t}, \ldots, y_{Pt} \right)^\prime$ represents the gene expression measurements at time $t$. In this work, we consider two related characterizations of a GRN: an adjacency matrix $G$, and a parameter matrix $\Theta$. For the former, let $G$ be a $P \times P$ matrix such that $G_{ij} = 1$ if gene $j$ regulates gene $i$, and $G_{ij} = 0$ otherwise. For the latter, we define $\Theta$ as the $P \times P$ parameter matrix of a GRN, where $\theta_{ij}$ represents the relationship between gene $j$ at time $t-1$ and gene $i$ at time $t$. For this matrix, a value of $\theta_{ij} = 0$ indicates that gene $j$ does not regulate gene $i$; if $\theta_{ij} > 0$ ($\theta_{ij} < 0$ respectively), gene $j$ activates (represses) gene $i$. Note that $P(\theta_{ij} = 0 \vert G_{ij} = 0) = 1$, and $P(\theta_{ij} = 0 \vert G_{ij} = 1) = 0$. We will further discuss the simultaneous use of the matrices $\Theta$ and $G$ in Section~\ref{subsec:proposal}. 

Our objective in this work is to determine which gene-to-gene interactions within the GRN may be inferred, based on their approximate posterior distributions. To accomplish this, we first introduce the Bayesian model used to model the time-course gene expression data $Y$ and corresponding gene regulatory network $\Theta$. After motivating the use of ABC methods in this context, we then introduce the ABC-MCMC algorithm of \citet{Marjoram2003} in greater detail, and describe our modifications for reverse engineering GRN.

\subsection{Bayesian Model}
\subsubsection{Likelihood specification}
For a given gene regulatory network $\Theta$, we model the time-course gene expression data as a time-homogeneous Markov model
\begin{equation}
Y \sim \prod_{t} f(\mathbf{y}_t; \mathbf{y}_{t-1}, \Theta).  \label{eqn:model}
\end{equation}
with the convention that $\mathbf{y}_0 = 0$. Several authors \citep[e.g.,][]{Beal2005, Opgen-Rhein2007, Wilkinson2009} have found that simple, linear models can in some cases yield good approximations of the dynamics occurring within complicated biological systems. To this end, one simple yet effective choice for the density $f$ in Equation~(\ref{eqn:model}) is a first-order vector autoregressive (VAR(1)) model:
\begin{equation}
\mathbf{y}_{t} = \Theta \mathbf{y}_{t-1} + \mathbf{e}_t \label{eqn:var1}
\end{equation} 
where $\mathbf{e}_t$ is an error term satisfying $E(\mathbf{e}_t) = \mathbf{0}$, $E(\mathbf{e}_t\mathbf{e}_t^\prime) = \Sigma$ (a $P\times P$ positive definite covariance matrix), and $E(\mathbf{e}_t\mathbf{e}_{t^\prime}^\prime) = 0$. In previous work \citep[e.g.,][]{Beal2005, Rau2010}, the errors $\mathbf{e}_t$ have additionally been assumed to follow a normal distribution, $\mathbf{e}_t \sim N(0, \Sigma)$. In this work, we do not impose any particular form for the distribution of the errors $\mathbf{e}_t$ beyond the assumptions on the first two moments previously mentioned.

\subsubsection{Network Prior Distributions}
To fully define the Bayesian model used for $Y$, we must also specify the prior distributions for the adjacency matrix $G$ and parameter matrix $\Theta$, $\pi(G)$ and $\pi(\Theta \vert G)$. In a GRN, as the number of genes ($P$) in a network increases, the number of possible interactions within the network quickly increases ($P\times P$). As a large number of genes may interact simultaneously with one another in very sophisticated regulatory circuits, the network topology itself may be quite complicated. Even so, certain properties of biological networks can be useful in limiting the support of the prior distribution to realistic network topologies. In particular, most genes are regulated just one step away from their regulator \citep{Alon2007}, and gene networks tend to be sparse, with a limited number of regulator genes \citep{Leclerc2008}. 

In keeping with these biological hypotheses, we elect to use uninformative prior distributions with some restrictions for both $\pi(G)$ and $\pi(\Theta \vert G)$. We restrict the number of regulators for each gene (referred to as the fan-in for each gene in the network). Because GRN are known to be sparse, we choose the prior on the adjacency matrix, $\pi(G)$, to be uniform over all possible structures, subject to a constraint on the maximum fan-in for each gene in the network, as has been suggested \citep{Friedman2000, Husmeier2003, Werhli2007}. This restriction is supported by the biological literature, as genes do not tend to be synchronously regulated by a large number of genes \citep{Leclerc2008}. For the parameter prior $\pi(\theta_{ij} \vert G_{ij} = 1)$, we use a uniform distribution, where the bounds are chosen to represent a realistic range of interaction magnitudes in GRN. In this work, we use bounds of -2 and 2 for all $\theta_{ij}$, as these correspond to strong repression and activation effects, respectively.

\subsection{ABC Motivation}
Given the likelihood and prior distributions defined in the previous section, our goal is to reverse engineer a GRN from observed expression data $Y$ via the posterior distribution
\begin{equation}
\pi(\Theta, G \vert Y) \propto f(Y \vert \Theta)\pi(\Theta\vert G)\pi(G). \nonumber 
\end{equation}
In some cases, the error term $\mathbf{e}_t$ included in the likelihood in Equation~(\ref{eqn:var1}) is assumed to follow a well-known distribution (e.g., a Normal distribution). This hypothesis would enable straightforward calculation of the likelihood, and in turn, the posterior distribution $\pi(\Theta\vert Y)$, whether through explicit calculation or a standard MCMC sampler. However, in this work we do not impose a specific distributional form for $\mathbf{e}_t$, and as such, the likelihood $f(Y \vert \Theta)$ cannot be evaluated. It is exactly in situations such as this that ABC methods have been successfully developed and applied in recent years. 

Simple ABC rejection methods (e.g., \citet{Pritchard1999}) have the advantage of being easy to code and generating independent observations, but can be extremely time consuming and inefficient, particularly in the case of GRN. To illustrate, we applied the following simple ABC rejection method to sample from the approximate posterior distribution $\pi(\Theta, G \vert Y)$:
\begin{enumerate}
\item Generate $G$ and $\Theta$ from $\pi(G)$ and $\pi(\Theta\vert G)$, respectively.
\item Generate one-step-ahead predictors $\mathbf{y}_t^\star$ from model \ref{eqn:var1}, given $\mathbf{y}_{t-1}$ and $\Theta^\star$ (see Section \ref{subsec:simulating} for a discussion of this simulation strategy).
\item Calculate the distance $\rho(Y,Y^\star)$ between $Y$ and $Y^\star$.
\item Accept $(\Theta^\star, G^\star)$ if $\rho \leq \epsilon$, where $\epsilon$ is chosen as described in Section~\ref{subsec:implementation}.
\end{enumerate}
Using this algorithm, only 5 proposed networks $(\Theta^\star, G^\star)$ are accepted out of a total of $1\times 10^7$ proposals (data not shown). Because such an approach is both inefficient and unpractical, we focus instead on the ABC-MCMC approach of \citet{Marjoram2003}.

\subsection{ABC Markov Chain Monte Carlo for Networks}
%	- Basics of ABC-MCMC method
The ABC-MCMC algorithm \citep{Marjoram2003} makes use of the standard Metropolis-Hastings scheme \citep{Hastings1970} to obtain samples from the approximate posterior distribution $\pi(\Theta, G \vert \rho(Y^\star, Y) < \epsilon)$. To accomplish this, matrices $\Theta^\star$ and $G^\star$ are proposed based on a proposal distribution $q(\cdot\vert\cdot)$ and subsequently used to simulate data $Y^\star$ based on a given model $f(\cdot \vert \Theta^\star)$. Simulated and observed data are compared using a distance function $\rho(\cdot)$ and tolerance $\epsilon$, and proposed parameters are accepted with probability 
\begin{equation*}
\alpha = \min \left \{ 1, \frac{\pi(\Theta^\star, G^\star)q(\Theta^i, G^i \vert \Theta^\star, G^\star)}{\pi(\Theta^i, G^i)q(\Theta^\star, G^\star \vert\Theta^i, G^i)} \mathbf{1}\left( \rho(Y^\star, Y) < \epsilon \right)\right \} \nonumber
\end{equation*}
where $\mathbf{1}(\cdot)$ is an indicator function that replaces the likelihood, and $\pi(\cdot)$ represents the prior distributions of $(\Theta, G)$. Under suitable regularity conditions \citep{Marjoram2003}, it is straightforward to show that the stationary distribution of the chain is indeed the approximate posterior distribution. If $\epsilon$ is sufficiently small, then this distribution will be a good approximation to the true posterior distribution $\pi(\Theta, G\vert Y)$. However, a balance must be achieved between a small enough tolerance to obtain a good approximation to the posterior and a large enough tolerance to allow for feasible computation time. \citet{Bortot2007} proposed a further adaptation of ABC-MCMC for the purpose of improving its mixing properties using data augmentation techniques, known as the ABC-MCMC augmented algorithm. Specifically, the parameter space is augmented with the tolerance $\epsilon$, which is treated as a model parameter with its own pseudo-prior distribution. Although this algorithm alleviates the problem of insufficient mixing, since larger values of $\epsilon$ may be accepted, it typically requires a much larger number of iterations than the original ABC-MCMC algorithm.

%	- Lead in to what needs to change (simulating data, proposal distributions, prior distributions)
Adapting the ABC-MCMC algorithm of \citet{Marjoram2003} to the context of GRN requires two important considerations to be taken into account: 1) computationally efficient methods for simulating data $Y^\star$ from a known GRN (defined by its parameter matrix $\Theta^\star$), and 2) an appropriate proposal distribution $q(\cdot \vert \cdot)$ for both the network structure and parameters. We refer to the algorithm incorporating these adaptations as the ABC for Networks (ABC-Net) method. For clarity, although we limit this discussion to data with a single biological replicate, the extension to multiple replicates is straightforward.
% The algorithm presented here in the case of a simple VAR(1) model but could easily be extended to much more complex models.

\subsection{Simulating Data for Gene Networks within ABC}
\label{subsec:simulating}
One of the most important considerations in adapting the ABC-MCMC algorithm to the inference of GRN is identifying an efficient simulator for proposed network parameter matrix $\Theta^\star$. Broadly, we simulate gene expression at time $t$ as a function of gene expression at the previous time point and the proposed parameter matrix $\Theta^\star$ using a VAR(1) model as in Equation \ref{eqn:var1}. Specifically, after setting $\mathbf{y}^\star_1 = \mathbf{y}_1$, we exploit the Markov property of the VAR(1) model to obtain one-step-ahead predictions (i.e., fitted values) of gene expression at time points $t = 2, \ldots, T$:
\begin{equation}
\mathbf{y}_t^\star = \Theta^\star \mathbf{y}_{t-1}. \label{eqn:varsimulator}
\end{equation}
Note that the one-step-ahead predictions for $\mathbf{y}_t$ are made using the observed data $\mathbf{y}_{t-1}$, and not the simulated data $\mathbf{y}_{t-1}^\star$. That is, we simulate data deterministically by calculating the expected value of gene expression at each time point given the network structure and observed expression values at the previous time point, rather than incorporating an estimate of noise in the simulated data. 

We are aware that the deterministic simulation procedure discussed above is somewhat unconventional in the ABC literature, primarily since it does not incorporate an estimate of noise in the simulated data $Y^\star$. More classically, repeated sampling is used to control the variability of the data by simulating several noisy datasets $\lbrace Y_1^\star, \ldots, Y_M^\star\rbrace$ for a given network $\Theta^\star$, with $M > 1$. Keeping this in mind, adding no noise can be seen as the limiting case of $M > 1$ replications of the dataset generated from the same $\Theta$, as advocated in some ABC procedures \citep{DelMoral2009}. In our case, the choice to use the one-step-ahead predictors as in Equation~(\ref{eqn:varsimulator}) is a practical one. More specifically, because the time-series expression data are modeled as a VAR(1) process, we found that adding noise at early time points simply had the effect of inducing wide discrepancies at later time points, as incorrect error terms compounded throughout the simulated time series. This had the effect of creating large distances $\rho(Y,Y^\star)$, even when the true network $\Theta$ was used to generate $Y^\star$.

Finally, the appropriateness of using a VAR(1) simulator, Equation (\ref{eqn:varsimulator}), is largely dependent on the noise present in observed data, as well as the adequacy of the assumption of time-invariant, first-order autoregressive dynamics for complicated GRN. In the absence of more detailed information about the underlying network, it may be reasonable to use a simple model such as the VAR(1) to generate simulated data. We note that the ABC-Net algorithm has the flexibility to incorporate arbitrary models as data simulators, provided they are computationally efficient. For instance, in some cases second-order models, nonlinear models, linear differential equations, draws from a Dirichlet process, or Michaelis-Menten kinetics may more aptly describe the dynamics of a particular GRN; in these cases, the appropriate simulator model would be used in place of Equation (\ref{eqn:varsimulator}).

\subsection{Two-Step Network Proposal Distributions}
\label{subsec:proposal}
%% Improve the computational efficiency of the algorithm (as in a PX algorithm)
Another important consideration is the proposal distribution $q(\cdot \vert \cdot)$ that defines the transition from the current proposal for a GRN to an updated proposal. Based on the current values of $G$ and $\Theta$, a two-step proposal distribution is used to produce new samples $G^\star$ and $\Theta^\star$ for the adjacency and parameter matrices, respectively. In this context, the adjacency matrix $G$ may be viewed as an auxiliary variable \citep{Damien1999}, which is introduced to simplify the Markov chain Monte Carlo algorithm. As such, the joint distribution of $G$ and $\Theta$ may be seen as a completion of the marginal density of $\Theta$ \citep{Robert2004} which facilitates simulation within the MCMC algorithm.  

In the first step, one of three basic moves \citep{Husmeier2005} is applied to the current adjacency matrix $G^i$: adding an interaction (i.e., changing a 0 to a 1), deleting an interaction (i.e., changing a 1 to a 0), or reversing the direction of an interaction (i.e., if $G_{ij} = 1$ and $G_{ji} = 0$, exchanging these two values). If $\mathcal{N}(G)$ represents the neighborhood size of a particular adjacency matrix $G$, (i.e., the number of other network structures that can be obtained by applying one of these three basic moves), the transition probability of the first step is given by $q(G^\star \vert G^i) = 1/ \mathcal{N}(G^i)$. 

\begin{figure}[t!]
\centering
\includegraphics[width = .5\textwidth, clip = true, trim = .1cm .1cm 1.75cm 18cm]{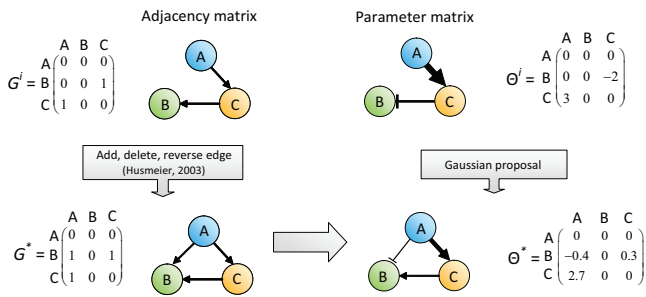}
\caption{Example of two-step proposal distribution for GRN. Top row: A network in iteration $i$ of the ABC-Net algorithm may be characterized both by its adjacency matrix $G^i$ (left) and its parameter matrix $\Theta^i$ (right). The former encodes only the presence (1) or absence (0) of an interaction. The latter encodes additional information about the magnitude of a particular interaction, where zeros indicate that an interaction is not present, positive values indicate an activation, and negative values indicate a repression (interactions with values further away from zero correspond to stronger effects). Bottom row: An updated network is proposed by adding, deleting, or reversing an interaction in $G^i$ to produce $G^\star$ (left). The parameter matrix $\Theta^i$ is updated using a Gaussian proposal distribution for the nonzero interactions of $G^\star$ to produce $\Theta^\star$ (right). Image taken from \citet{Rauthesis2010}\label{fig:netparam}}
\end{figure}

In the second step, the proposal distribution of $\Theta$, given the current value $\Theta^i$ and the updated adjacency matrix $G^\star$, is defined to be
\begin{equation}
q(\theta_{ij} \vert \theta_{ij}^i, G_{ij}^\star) \sim \left\{ 
\begin{array}{l l}
  0 & \quad \mbox{if $G_{ij}^\star = 0$}\\
  N(\theta_{ij}^i, \sigma_\Theta^2) & \quad \mbox{if $G_{ij}^\star \ne 0$}\\ \end{array} \right. \label{eqn:Gaussproposal}
\end{equation}
where $\sigma_\Theta^2$ is the variance of the proposal distribution, and $\sigma_\Theta$ may be tuned to obtain an empirical acceptance rate between 15\% and 50\%, as recommended in \citet{Gilks1996}. A simple example of the two-step proposal distribution for GRN is shown in Figure \ref{fig:netparam}.

It is worth noting that the introduction of the adjacency matrix $G$ is not strictly necessary to accomplish the two-step proposal described above. For instance, it would be straightforward to define the target with respect to a mixture of singular measures, e.g., a dirac mass and a Gaussian density \citep[see][for more details]{Gottardo2004}. Furthermore, the three proposal moves (add, delete, and reverse a network edge) could be defined using a mixture of kernels that include selection probabilities depending on the current state. Our primary motivation for including $G$ is based on the approach for learning Bayesian networks in Chapter 2 of \citet{Husmeier2005}, which clearly distinguishes the network structure (i.e., the set of edges and nodes represented by the adjacency matrix $G$) from the network parameters (the matrix $\Theta$). The three-move proposal strategy we apply for the structure of the network is based on this intuitive representation, and is rather popular in Bioinformatics \citep[see e.g.,][]{Husmeier2005}.

\subsection{ABC-Net Implementation}
\label{subsec:implementation}

The output from the ABC-Net algorithm consists of dependent samples from the stationary distribution of the chain, $f(\Theta, G \vert \rho(Y^\star, Y) \leq \epsilon)$. In practice, because saving all iterations from the MCMC run can take up a large amount of storage (particularly as the size of the network increases) and consecutive draws tend to be highly correlated, we thin the chain at every $50^\mathrm{th}$ iteration. Additionally, as with many MCMC methods, a burn-in period is implemented to reduce the impact of initial values and to improve mixing for the chain. The length $b$ of the burn-in depends on the starting values of the chain, $\Theta^0$ and $G^0$, the rate of convergence of the chain, and the similarity of the transition mechanism of the chain to the approximate posterior distribution. We follow the suggestion of \citet{Geyer1992}, setting $b$ to between 1\% and 2\% of the run length $n$. 

We also implement a ``cooling" procedure during the burn-in period similar to that used in \citet{Ratmann2007}, where acceptance of $(G^\star, \Theta^\star)$ is controlled by a decreasing sequence of thresholds, until the minimum pre-set value $\epsilon$ is reached. Note that tempering the acceptance threshold $\epsilon$ in this way reduces the number of accepted parameters as the number of iterations increases. This cooling scheme also addresses the poor mixing often observed in the ABC-MCMC algorithm, as larger tolerances in the early iterations of the burn-in are associated with higher acceptance rates. A total of 200 iterations are run for each of ten cooled threshold values, and the burn-in period is repeated if the empirical acceptance rate is less than 1\%. This ensures a minimum burn-in period of 2000 iterations, with additional iterations included for chains affected by poor mixing. 

Because the ABC-Net algorithm relies on a comparison between simulated and observed data to avoid a likelihood calculation, long chains are required to ensure the adequacy of the approximation. Although a single long chain could be run, it is also possible to run multiple overdispersed chains. In practice, we run 10 independent chains of length $1\times 10^6$ simultaneously (rather than a single chain of length $1 \times 10^7$). This approach contributes a two-fold benefit, as calculations can be performed in parallel to improve computational speed and a convergence assessment can be conducted using the Gelman-Rubin statistic $R$ \citep{Gelman1992}. Following the recommendation in \citet{Gilks1996} we declare chain convergence if $\hat{R} < 1.2$ for all parameters in $\Theta$. After the chains have converged, draws corresponding to the smallest 1\% of the distance criterion are retained for inference. 

%%%%%%%%%%%%%%%%%%%%%%%%%%%%%%%%%%%%%%%%%%%%%%%%%%%%%%%%%%%%%%%%%%%%%%%%%%%%%%%%%%%%%%%%%%%%%%%%%%%%%%%%%%%%%%
%%%%%%%%%%%%%%%%%%%%%%%%%%%%%%%%%%%%%%%%%%%%%%%%%%%%%%%%%%%%%%%%%%%%%%%%%%%%%%%%%%%%%%%%%%%%%%%%%%%%%%%%%%%%%%

\section{Simulation Study Based on the Raf Pathway}
\label{s:sims}
In this simulation study, we focus on four specific aspects related to the performance of the ABC-Net: the distance function $\rho$ and tolerance $\epsilon$, the sensitivity to prior distribution bounds, the suitability of the model used to generate simulated data when more complicated dynamics are at play, and the effect of increasing the amount of noise present in the observed data. To do so, we focus on the Area Under the Curve (AUC) of the Receiver Operating Characteristic (ROC) curve as an indicator of performance, as well as qualitative examinations of the approximate posterior distributions of interactions in the network. To calculate the AUC , we retain only the samples corresponding to the smallest 1\% of distances $\rho(Y^\star, Y)$ for inference. Based on these samples, we calculate the bounds of the $\alpha\%$ credible intervals for each gene-to-gene interaction, where $\alpha = \lbrace 1, \ldots, 100 \rbrace$. If the $\alpha\%$ credible interval for a particular interaction does not contain 0, the gene-to-gene interaction is declared to be present; otherwise, the interaction is declared to be absent. In this way, because the simulation setting determines which interactions are truly present and absent, true positives, false positives, true negatives, and false negatives may be calculated for each $\alpha$, and the AUC may subsequently be calculated. The values of $\alpha$ may be adjusted for multiple testing, if necessary. 

\begin{figure}[t!]
\label{RafpathwayDiagram}
\centering
\includegraphics[height = .5\textwidth, angle = -90, clip = true, trim = 2cm 3cm 9cm 3cm]{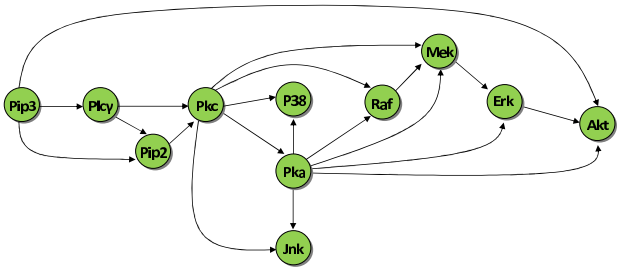}
\caption{The currently accepted gold-standard Raf signalling pathway \citep{Werhli2007}, which describes the interactions of eleven phosphorylated proteins in primary human immune system cells \citep{Sachs2005}. Nodes represent the proxy genes of each of the eleven proteins (i.e., the genes that are transcribed and translated into the corresponding proteins), and arrows indicate the direction of signal transduction. Image taken from \citet{Rauthesis2010}\label{Rafpathway}}
\end{figure}

\subsection{Simulation Design}
Rather than defining an arbitrary network $\Theta$, we instead make use the structure of a well-characterized pathway in human immune system cells involving the Raf signalling protein \citep{Sachs2005}. We generate data based on 11 genes, where the adjacency matrix $G^{\mathrm{Raf}}$ is defined using the structure of the currently accepted Raf signalling network (Figure \ref{RafpathwayDiagram}). If an interaction is present from gene $j$ to gene $i$, we sample $\theta_{ij}^{\mathrm{Raf}}$ uniformly from the interval $(-2,-0.25) \cup (0.25, 2)$, and otherwise $\theta_{ij}^{\mathrm{Raf}} = 0$. The bounds for non-zero gene-to-gene interactions were chosen to represent a range of moderate to strong interactions among genes. We generate one replicate of expression data for each of the 11 genes over 20 time points, using the VAR(1) model
\begin{equation}
\mathbf{y}_t = \Theta^{\mathrm{Raf}}\mathbf{y}_{t-1} + \mathbf{z}_t \label{eqn:simsRafeqn}
\end{equation}
for $t = 1, ..., T$, where $\mathbf{y}_1 \sim N(0,I)$, and $\mathbf{z}_t \sim N(0, \sigma^2)$. For each simulation, unless otherwise noted, the noise standard deviation is set to $\sigma = 1$, the Gaussian proposal standard deviation in Equation (\ref{eqn:Gaussproposal}) is set to $\sigma_\Theta = 0.5$, and the maximum fan-in is constrained to 5 or less. 

\subsection{Choice of $\rho$ and $\epsilon$} %Reference some other studies...
\label{sssec:dist}
The distance function $\rho$ and threshold $\epsilon$ are essential components to the ABC-Net method, as they directly affect the probability that simulated data $Y^\star$ generated by a network $\Theta^\star$ are accepted as being ``close enough" to the observed data. Although there are many potential options for this distance function, we focus on a comparison among the Manhattan, Euclidean, Canberra, and Multivariate Time-Series \citep[MVT;][]{Lund2009} distances (see Appendix). For each choice of $\rho$, we propose a heuristic method where 5000 randomly generated networks are used to simulate data, and the corresponding distances $\rho(Y^\star, Y)$ are calculated for each. Subsequently, $\epsilon$ is set to be either the 1\%, 5\%, or 10\% quantile of these distances associated with 5000 randomly generated networks. The number of randomly generated networks was chosen based on a set of preliminary simulations that indicated that the quantiles for the corresponding distances $\rho(Y^\star, Y)$ seemed to stabilize for 5000 or more networks (data not shown). For larger networks, further exploratory simulations may need to be performed to ensure that this number is not too small. Each combination of $\rho$ and $\epsilon$ was repeated over five independent datasets in order to include an assessment of their variability (only two datasets were simulated for the MVT distance due to its computational burden). 

\begin{figure}[t!]
\label{fig:distanceAUCs}
\centering
\includegraphics[width = 0.5\textwidth]{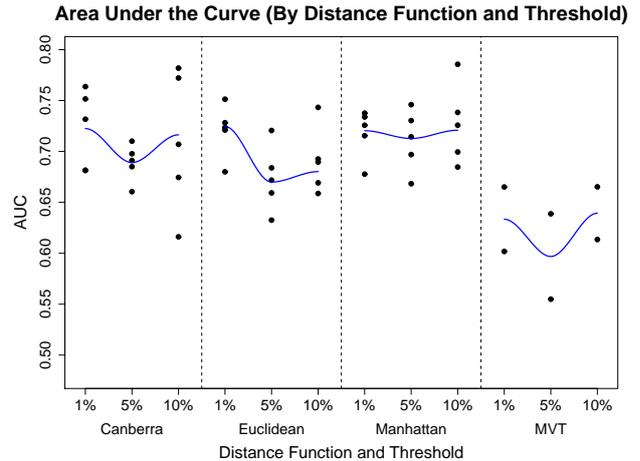}
\caption{Area Under the Curve (AUC) of the Receiver Operating Characteristic (ROC) curve for four choices of distance functions in the ABC-Net algorithm: Canberra, Euclidean, Manhattan, and MVT distances. Black dots represent the value of the AUC for each of five independent datasets per threshold and distance function (with the exception of the MVT distance, which was limited to two datasets due to its computational burden). The threshold $\epsilon$ was set at the 1\%, 5\%, and 10\% quantiles from 5000 randomly generated networks. Blue lines represent loess curves \citep{Cleveland1979}. Image taken from \citet{Rauthesis2010}\label{fig:distanceAUCs}}
\end{figure}

Each distance function under consideration calculates and penalizes differences between simulated and observed data in a different way. In particular, the behavior of the MVT function appears to differ from that of the other distance functions, with much lower AUC values for each combination of $\rho$ and $\epsilon$ (Figure \ref{fig:distanceAUCs}). The Canberra, Euclidean, and Manhattan distances all appear to be on par with one another, particularly when $\epsilon$ is set at the 1\% quantile of distances. However, based on the criterion of AUC alone, there does not seem to be strong evidence that favors one choice among the Canberra, Euclidean, and Manhattan distances, particularly for a cutoff of $\epsilon$ = 1\%. That is, although the MVT distance is a poor choice of distance function within the ABC-Net algorithm, the remaining distances yield similar results. Because it enjoys a slight advantage over the Manhattan and Canberra distances in terms of computation time, we use the Euclidean distance with $\epsilon$ set to the 1\% threshold for the remainder of the simulations.

\begin{figure}[t!]
\label{fig:GRpriorbounds}
\centering
\includegraphics[width = .5\textwidth]{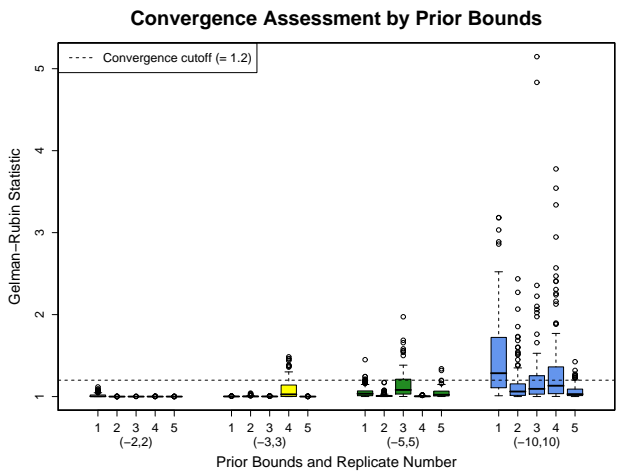}
\caption{Gelman-Rubin statistics ($\hat{R}$) for each replicate of four choices of bounds on the prior distribution $\pi(\Theta\vert G)$. The black dotted line indicates a value of $\hat{R} = 1.2$, the cutoff at which convergence is declared among ten independent chains in the ABC-Net algorithm. Image taken from \citet{Rauthesis2010}\label{fig:priorboundsgr}}
\end{figure}

\subsection{Sensitivity to prior distribution bounds}
Although the prior bounds (-2 and 2) for $\pi(\Theta \vert G)$ are reasonable for the context of GRN, we also consider the following bounds: (-3,3), (-5,5), and (-10,10). These intervals include somewhat ``unrealistic" values for $\Theta$, but are more diffuse (and hence less informative). The greatest effect of using less informative prior distributions is in terms of the convergence of the ten independent chains, as assessed by the Gelman-Rubin statistic (Figure \ref{fig:GRpriorbounds}). This is most evident for prior bounds of (-10,10), where a large number of interactions exceed the convergence cutoff of 1.2 by a large amount. It is perhaps unsurprising that wider prior bounds lead to problems in chain mixing and convergence, and thus highlights the need for well-chosen prior bounds for the inference of GRN. 

\begin{figure*}[ht!]
\centering
\includegraphics[width = .7\textwidth, clip = true, trim = 0cm 0 3cm 10cm]{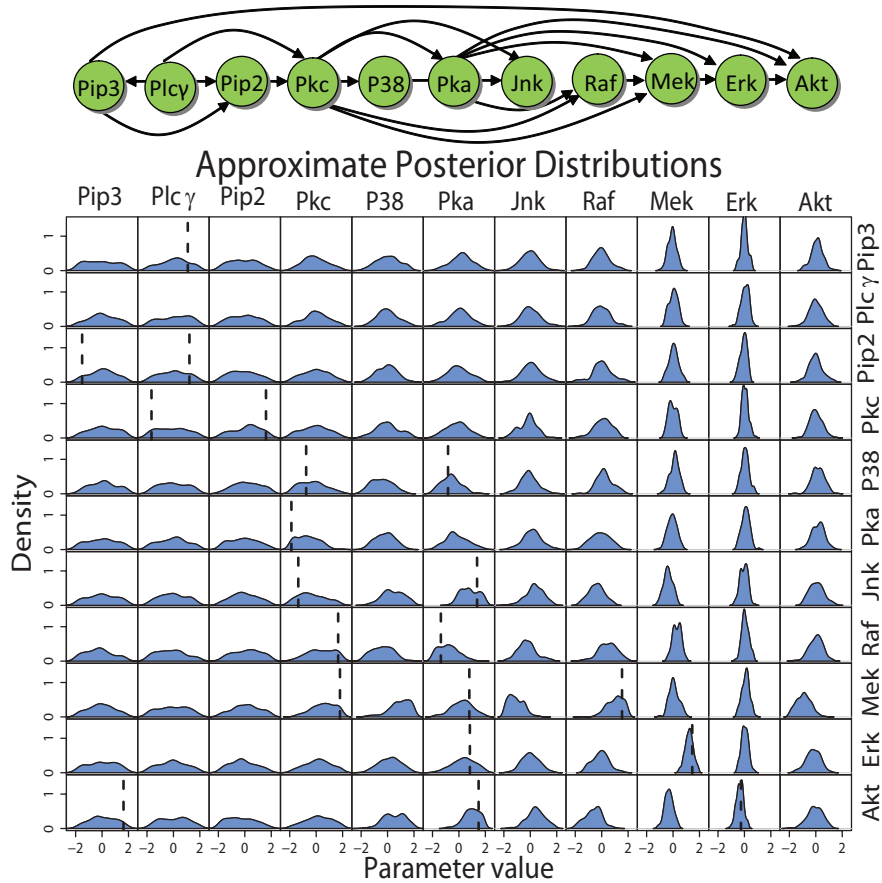}
\caption{The structure of the true Raf signalling pathway, $\Theta^\mathrm{Raf}$, and a graphical matrix of the marginal approximate posterior distributions for every interaction in the network, with prior bounds (-2,2). Each element of the graphical matrix corresponds to the same element of $\Theta^\mathrm{Raf}$, i.e., the density in the second row and first column corresponds to $\theta_{21}^\mathrm{Raf}$ (Pip3$\rightarrow$Plc$\gamma$). The x-axis of each plot represents the values of each parameter $\theta_{ij}^\mathrm{Raf}$, and the y-axis represents the corresponding density. Black dotted lines are included on plots where $\theta_{ij}^\mathrm{Raf} \ne 0$ at the true value. Image taken from \citet{Rauthesis2010}\label{fig:boundspost2}}
\end{figure*}

We also consider the effect of the choice of prior bounds on the shape of the approximate posterior distributions in the network. In Figure \ref{fig:boundspost2}, a graphical matrix of the marginal approximate posterior distributions of each interaction in the network is given for prior bounds (-2,2). As may be expected, the approximate posteriors are generally more diffuse when wider prior bounds are used. However, regardless of the choice of prior bound, some gene-to-gene interactions consistently have very flat (diffuse) approximate posterior distributions (e.g., those in the Pip3 column), while others tend to be consistently peaked (e.g., those in the Erk column). We refer to gene-to-gene interactions with these two characteristics as ``flexible" and ``rigid", respectively. Interestingly, in this simulation, the most rigid interactions appear to correspond to regulators that are furthest downstream in the simulated pathway (Mek, Erk, and Akt), while those furthest upstream appear to be the most flexible. In the context of the ABC-Net method, this suggests that rigid interactions (e.g., Mek$\rightarrow$Erk) in $\Theta^\star$ must take on values within a tight interval in order to generate simulated data $Y^\star$ that are close (in terms of $\rho$ and $\epsilon$) to the observed data $Y$. Conversely, flexible interactions (e.g., Pip3$\rightarrow$Pip3) can take on values within a much wider interval without negatively affecting the proximity of simulated and observed data. Thus, it is likely that the model is most sensitive to parameters with narrow credible intervals (rigid interactions) and least sensitive to those that cannot accurately be localized (flexible interactions) by the approximate posterior distribution \citep{Toni2009}. That is, it appears that some interactions may intrinsically be easy to infer even with relatively wide prior bounds, while others cannot be accurately determined even with reasonable prior distribution bounds.

\begin{table}
\centering
\caption{Alternative models used to generate observed data $Y$: an ordinary differential equation (ODE), a second-order VAR model (VAR(2)), a first-order nonlinear VAR model (VAR-NL(1)), and a second-order nonlinear VAR model (VAR-NL(2))\label{tab:othermodels}}
\begin{tabular}{c|l}
\hline
Model & Network Equations to Generate $Y$\\ 
\hline
\multirow{2}{*}{VAR-NL(1)} 
& $\mathbf{y}_1 = \mathbf{z}_1$ \\
& $\mathbf{y}_t = \Theta_1^\mathrm{Raf} \mathbf{y}_{t-1}^{-1} + \mathbf{z}_t$, \\
&\hspace{.6cm}for $t = 2, \ldots, T$\\
& $\mathbf{z}_t \sim N(0,1)$ for $t = 1, \ldots, T$\\
\hline 
\multirow{3}{*}{VAR(2)} 
& $\mathbf{y}_1 = \mathbf{z}_1$ \\
& $\mathbf{y}_2 = \Theta_1^\mathrm{Raf} \mathbf{y}_1 + \mathbf{z}_2 $\\
& $\mathbf{y}_t = \Theta_1^\mathrm{Raf} \mathbf{y}_{t-1} + \Theta_2^\mathrm{Raf} \mathbf{y}_{t-2} + \mathbf{z}_t$, \\
&\hspace{.6cm}for $t = 3, \ldots, T$\\
& $\mathbf{z}_t \sim N(0,1)$ for $t = 1, \ldots, T$\\
\hline
\multirow{2}{*}{VAR-NL(2)} 
& $\mathbf{y}_1 = \mathbf{z}_1$ \\
& $\mathbf{y}_2 = \Theta_1^\mathrm{Raf} \mathbf{y}_{t-1}^{-1} + \mathbf{z}_2$\\
& $\mathbf{y}_t = \Theta_1^\mathrm{Raf} \mathbf{y}_{t-1}^{-1} + \Theta_2^\mathrm{Raf} \mathbf{y}_{t-2} + \mathbf{z}_t$,\\
&\hspace{.6cm}for $t = 3, \ldots, T$\\
& $\mathbf{z}_t \sim N(0,1)$ for $t = 1, \ldots, T$\\
\hline
\multirow{11}{*}{ODE}
& $y_\mathrm{Pkc}^\prime = 0.18y_{\mathrm{Plc}\gamma} - 0.75y_\mathrm{Pip2}$\\
& $y_\mathrm{Raf}^\prime = -0.28y_\mathrm{Pkc} + 0.62y_\mathrm{Pka}$\\
& $y_\mathrm{Mek}^\prime = 0.63y_\mathrm{Pkc} - 0.97y_\mathrm{Raf} - 0.52y_\mathrm{Pka}$\\
& $y_\mathrm{Erk}^\prime = 0.70y_\mathrm{Mek} - 0.94y_\mathrm{Pka}$\\
& $y_\mathrm{Pka}^\prime = 0.31y_\mathrm{Pkc}$\\
& $y_\mathrm{Akt}^\prime = 0.28y_\mathrm{Erk} + 0.60y_\mathrm{Pka} + 0.92y_\mathrm{Pip3}$\\
& $y_\mathrm{P38}^\prime = -0.19y_\mathrm{Pkc} - 0.32y_\mathrm{Pka}$\\
& $y_\mathrm{Jnk}^\prime = 0.24y_\mathrm{Pkc} + 0.98y_\mathrm{Pka}$\\
& $y_{\mathrm{Plc}\gamma}^\prime = 0$\\
& $y_\mathrm{Pip3}^\prime = -0.28y_{\mathrm{Plc}\gamma}$\\
& $y_\mathrm{Pip2}^\prime = 0.83y_{\mathrm{Plc}\gamma} - 0.98y_\mathrm{Pip3}$\\
\hline
\end{tabular} 
\end{table}

\subsection{Suitability of VAR(1) Simulator}
The applicability of the ABC-Net method to real GRN relies heavily on its ability to accurately simulate data for a given network structure. It is feasible that real biological systems do not follow a VAR(1) model, and in fact, that they arise from very complicated, nonlinear relationships. To assess how the ABC-Net method performs when observed data $Y$ are actually generated from more complicated models, we focus on four models (Table \ref{tab:othermodels}): a first-order nonlinear VAR model (VAR-NL(1)), a second-order VAR model (VAR(2)), a second-order nonlinear VAR model (VAR-NL(2)), and an ordinary differential equation (ODE). For the VAR models, $\Theta_1^\mathrm{Raf}$ and $\Theta_2^\mathrm{Raf}$ were each defined using the structure of the Raf signalling network, where existing interactions were sampled uniformly from the interval $(-2,-0.25) \cup (0.25, 2)$ and otherwise set to 0. For the ODE model, coefficients were randomly drawn from a $\mathcal{U}(-1,1)$ distribution and initial values for all genes were set to 1. After solving the ordinary differential equations for time points $t = 1, \ldots, 20$, random noise sampled from $N(0,1)$ was added to each measurement at each time point. 

\begin{figure}[t!]
\centering
\includegraphics[width = .5\textwidth]{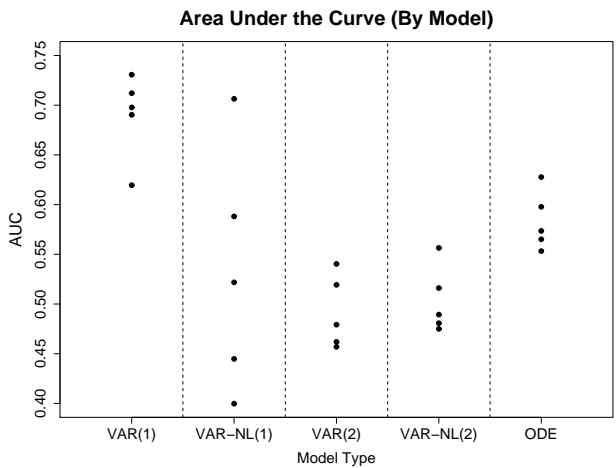}
\caption{Area Under the Curve (AUC) of the Receiver Operating Characteristic (ROC) curve for five different model choices to generate $Y$: VAR(1), VAR-NL(1), VAR(2), VAR-NL(2), and ODE. Black dots represent the value of the AUC for each of five independent datasets per bound. Image taken from \citet{Rauthesis2010}\label{fig:aucresults_other}}
\end{figure}

It is not surprising that the ABC-Net has the best performance in terms of AUC for the VAR(1) model, as the data $Y$ are generated with the same model that is used to simulate $Y^\star$ (Figure \ref{fig:aucresults_other}). For the other simulator models, the performance of the algorithm noticeably declines, with the lowest AUC values observed for the two second-order models, VAR(2) and VAR-NL(2). The nonlinear first-order VAR model shows wide variability in its results, ranging from an AUC of just over 0.40 to over 0.70. Of the alternative models, the ordinary differential equation appears to have the highest performance in terms of AUC. As a final note concerning the performance of the ABC-Net algorithm when alternative models are used to generate $Y$, recall that the simulator described in Section \ref{subsec:simulating} has the flexibility to incorporate alternative models, provided they are computationally efficient. In this respect, the VAR(1) model may be viewed as a kind of robust null model to apply when nothing is precisely known about the dynamics of a particular system. However, in cases where other models are known to better fit a given set of data (e.g., a second order or non-linear model), the ABC-Net method can be adapted accordingly.

\begin{figure}[t!]
\centering
\includegraphics[width = .5\textwidth]{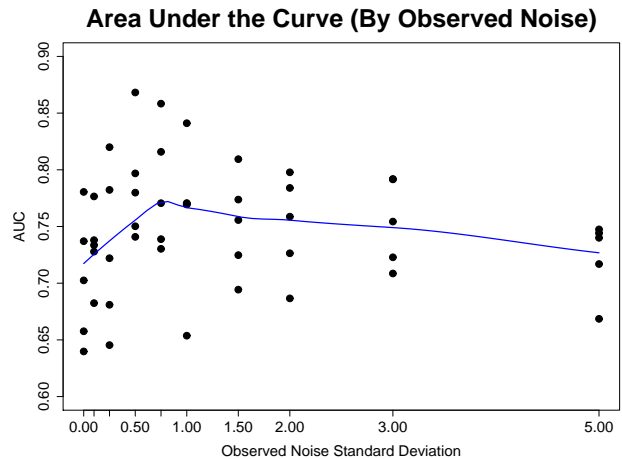}
\caption{Scatterplots of the Area Under the Curve (AUC) of the Receiver Operating Characteristic (ROC) curve for the ABC-Net algorithm, with differing values of noise standard deviation $\sigma$ (0, 0.1, 0.25, 0.5, 0.75, 1, 1.5, 2, 3, 5). Five datasets were generated for each value of noise standard deviation. The blue line represents a loess curve \citep{Cleveland1979}. Image taken from \citet{Rauthesis2010}\label{fig:AUCplotdistance}}
\end{figure}

\subsection{Effect of Noise in Observed Data}
We expect that increasing amounts of noise in the observed data (i.e., $\sigma$ in Equation 
\ref{eqn:simsRafeqn}) lead to reduced performance for the ABC-Net algorithm, particularly since the VAR(1) simulator uses one-step ahead predictors to simulate data based on a given network $\Theta^\star$. To evaluate this, we consider 
\begin{equation*}
\sigma = \lbrace 0, 0.1, 0.25, 0.5, 0.75, 1, 1.5, 2, 3, 5\rbrace,
\end{equation*} where $\mathbf{z}_t \sim N(0, \sigma)$. The AUC results (Figure \ref{fig:AUCplotdistance}) indicate that the presence of increasing noise over the investigated range does seem to negatively affect the performance of the ABC-Net algorithm, although only for relatively large values of $\sigma$ (e.g., $\sigma$ = 5). As the noise standard deviation increases, it is not surprising that the performance of the algorithm deteriorates, since the one-step-ahead predictors fall increasingly further from the observed data (even when the true network is used).

\begin{figure*}[ht!]
\centering
\includegraphics[width = .68\textwidth, clip = true, trim = 0 0 3cm 9cm]{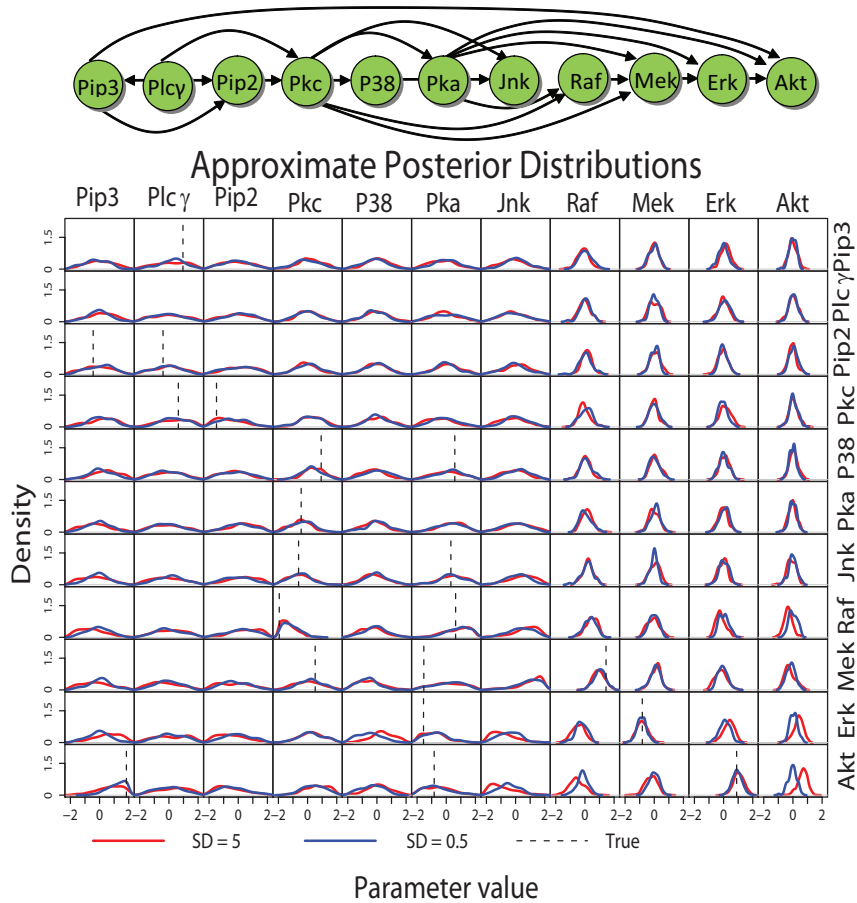}
\caption{The structure of the true Raf signalling pathway, $\Theta^\mathrm{Raf}$, and a graphical matrix of the marginal approximate posterior distributions for every interaction in the network, for $\sigma = 0.5$ and 5. Each element of the graphical matrix corresponds to the same element of $\Theta^\mathrm{Raf}$, i.e., the density in the second row and first column corresponds to $\theta_{21}^\mathrm{Raf}$ (Pip3$\rightarrow$Plc$\gamma$). The x-axis of each plot represents the values of each parameter $\theta_{ij}^\mathrm{Raf}$, and the y-axis represents the corresponding density. Red and blue lines correspond to results obtained with $\sigma = 5$ and $\sigma = 0.5$, respectively. Black dotted lines are included on plots where $\theta_{ij}^\mathrm{Raf} \ne 0$ at the true value. Image taken from \citet{Rauthesis2010}\label{fig:postmatrix_noise}}
\end{figure*}

We also examine the approximate posterior distributions of the network for two different values of noise standard deviation, $\sigma = 0.5$ and $\sigma = 5$ (Figure \ref{fig:postmatrix_noise}). For the most part, posterior distributions for both $\sigma = 0.5$ and $\sigma = 5$ seem to have the same general shape, with some occasional discrepancies (e.g., Akt$\rightarrow$Akt and Akt$\rightarrow$Erk). In addition, as in previous simulations, we note once again the marked difference in posterior distributions between rigid interactions (peaked distributions) and flexible interactions (diffuse distributions). Regardless of the amount of noise incorporated into the simulated data for the Raf signalling pathway, the approximate posterior distributions for the upstream and downstream portions of the network are consistently flexible and rigid, respectively. This seems to indicate that some interactions are intrinsically easier to infer (even in the presence of increased noise), while others cannot be accurately determined regardless of the amount of noise in the data. As such, the flexibility and rigidity of interactions in a given system likely plays an important role in the global inferability of the network structure. 

%%%%%%%%%%%%%%%%%%%%%%%%%%%%%%%%%%%%%%%%%%%%%%%%%%%%%%%%%%%%%%%%%%%%%%%%%%%%%%%%%%%%%%%%%%%%%%%%%%%%%%%%%%%%%%
%%%%%%%%%%%%%%%%%%%%%%%%%%%%%%%%%%%%%%%%%%%%%%%%%%%%%%%%%%%%%%%%%%%%%%%%%%%%%%%%%%%%%%%%%%%%%%%%%%%%%%%%%%%%%%
	
\section{Application to S.O.S. DNA Repair System in \textit{Escherichia coli}}
\label{s:app}

The S.O.S. DNA repair system in the \textit{Escherichia coli} bacterium is a well-known gene network that is responsible for repairing DNA after damage. The full network is made up of about thirty genes working at the transcriptional level.  The behavior of these genes in the presence of DNA damage has been well characterized \citep{Ronen2002}. Specifically, under normal conditions a master repressor called lexA represses the expression of the genes responsible for DNA repair. However, when one of the S.O.S proteins (recA) senses DNA damage by binding to single-stranded DNA, it becomes activated and provokes the autocleavage of lexA. The subsequent drop in the levels of lexA suspends the repression of the S.O.S. genes, and these genes become activated. Once DNA damage has been repaired, the level of recA drops, which allows lexA to reaccumulate in the cell and subsequently repress the S.O.S. genes. At this point, the cells return to their original state. Although the network itself is quite small, its simple structure allows the cell to react in very sophisticated ways to conditions within the cell.

\begin{figure*}[t!]
\centering
\includegraphics[width = .68\textwidth, clip = true, trim = 0 0 0 13.5cm]{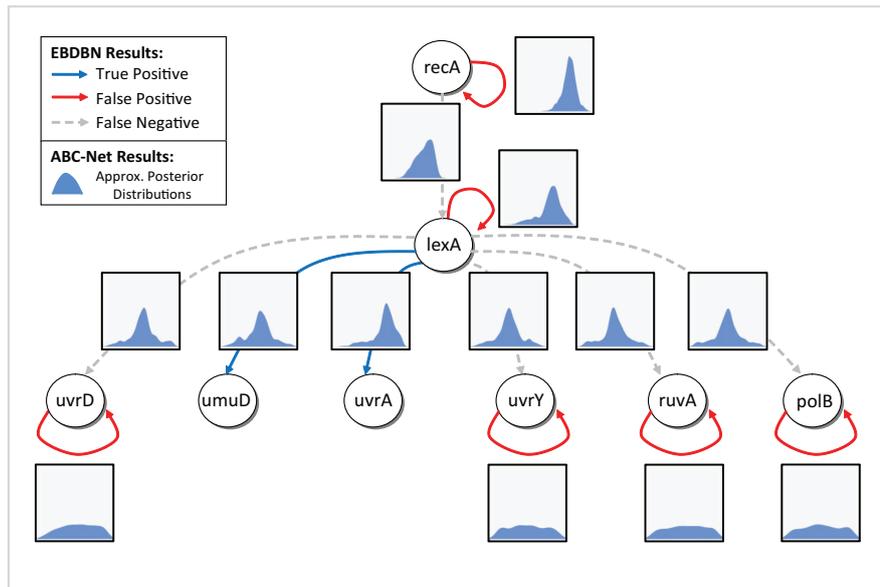}
\caption{Results for the S.O.S DNA repair system for the EBDBN and ABC-Net methods. Blue and red solid edges in the graph represent gene-to-gene interactions identified by the EBDBN method that are ``true positives" and ``false positives," according to the known behavior of genes in the S.O.S. network. Dotted gray lines represent gene-to-gene interactions supported by the literature that are not identified by the EBDBN method. Blue-filled densities represent the marginal approximate posterior distributions found through the ABC-Net method. The feedback loops on the S.O.S. genes (uvrD, uvrY, ruvA, and polB) appear to be flexible, while others exhibit greater rigidity. Image taken from \citet{Rauthesis2010}\label{fig:EcoliEBDBNABC}}
\end{figure*}

\subsection{Data}
We focus on a sub-network within the S.O.S. DNA repair system made up of eight genes: uvrD, lexA, umuD, recA, uvrA, uvrY, ruvA, and polB. Using green fluorescent protein (GFP) reporter plasmids, \citet{Ronen2002} measured the expression of these eight genes at fifty time points (every six minutes following ultraviolet irradiation of the cells to provoke DNA damage).  The quantity of GFP is proportional to the quantities of the corresponding S.O.S. proteins, which are in turn proportional to the corresponding mRNA production rates \citep{Perrin2003}. As such, it is reasonable to assume that the data of \citet{Ronen2002} directly indicate the expression levels of each of the S.O.S. genes. These data are available at the authors' website (\texttt{http://www.weizmann.ac.il/mcb/UriAlon}). In addition, the study performed by \citet{Ronen2002} consisted of two different experiments for each of two different intensities of ultraviolet light (Experiments 1 and 2 at 5 $Jm^{-2}$, and Experiments 3 and 4 at 20 $Jm^{-2}$). One recent study by \citet{Charbonnier2010} found that Experiments 1 and 4 systematically led to poor results for network inference methods, although nothing should distinguish them from the other two experiments. As such, we focus the rest of our discussion on the data collected in Experiment 3, which was measured with the higher level of ultraviolet light.

\subsection{Analysis}

In addition to the ABC-Net method, we apply the Empirical Bayes Dynamic Bayesian Network (EBDBN) approach of \citet{Rau2010} with a hidden state dimension of $K$ = 0, where a 99.9\% cutoff is used as a threshold for the z-scores of the gene-to-gene interactions. This particular method is chosen to illustrate the benefit of using the ABC-Net approach in tandem with other inference methods. As before, we set the Gaussian proposal standard deviation in Equation (\ref{eqn:Gaussproposal}) to $\sigma_\Theta = 0.5$, and we ran the ABC-Net method for ten independent chains of length $1\times 10^6$, with a thinning interval of 50. The VAR(1) simulator is used to generate simulated data $Y^\star$, and the prior bounds of $\pi(\Theta\vert G)$ are set to (-2,2). We used the Euclidean distance function, where the threshold $\epsilon$ is selected using the previously described heuristic method (Section \ref{sssec:dist}), based on the 1\% quantile of distances for 5000 random networks. Due to the small size of the network, the maximum fan-in was constrained to 2 or less.

The gene-to-gene interactions identified by the EBDBN method are illustrated in Figure \ref{fig:EcoliEBDBNABC}, where blue and red solid edges represent ``true positives" and ``false postives," according to the previously described behavior of the S.O.S. network. We use these terms somewhat loosely, because even for well-understood networks such as the S.O.S. DNA repair system, the absence of a particular gene-to-gene interaction in the literature cannot indicate with absolute certainty that such a relationship is absent. Gray dotted lines represent gene-to-gene interactions supported by the literature that are not identified by the EBDBN method. We also examine the marginal approximate posterior distributions for each of these interactions (Figure \ref{fig:EcoliEBDBNABC}), as obtained by the ABC-Net method. As previously seen in the simulation study, these posterior distributions seem to fall into two categories: flexible interactions (the feedback loops on uvrD, uvrY, ruvA, and polB) and rigid interactions (the others). That is, gene-to-gene interactions identified by the EBDBN with rigid approximate posterior distributions appear to be supported by substantial evidence, as those parameters are restricted to a smaller range of values in their posterior distributions. On the other hand, those associated with flexible approximate posterior distributions may indeed represent false positives, since those parameters take on a wider range of values without negatively impacting the proximity of simulated and observed data in the ABC-Net algorithm. In this way, the ABC-Net method can help yield complementary information about specific gene-to-gene interactions, as well as the overall dynamics of a given biological system. That is, the ABC-Net method can serve as a useful reference tool to confirm or belie results obtained by a more specific model. 

\begin{figure}[t!]
\centering
\includegraphics[width = .5\textwidth, clip = true, trim = 0 0 0 16cm]{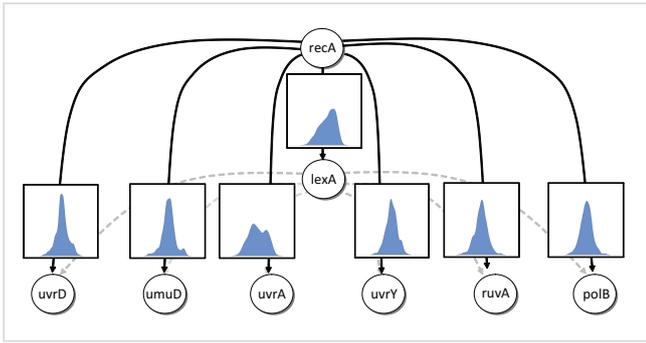}
\caption{Interactions exhibiting the highest rigidity in the S.O.S DNA repair system for the ABC-Net method. Dotted gray lines represent gene-to-gene interactions supported by the literature. Blue-filled densities represent the marginal approximate posterior distributions found through the ABC-Net method. The most rigid interactions in the network connect the recA protein directly to the S.O.S. genes, bypassing the lexA master regulator. Image taken from \citet{Rauthesis2010}\label{fig:EcoliABCrigid}}
\end{figure}

In addition to comparing the results of the EBDBN and ABC-Net methods, we also examine the most rigid approximate posterior distributions identified by the latter method (Figure \ref{fig:EcoliABCrigid}). Interestingly, all of the most rigid interactions in the S.O.S. DNA repair system are those directly connecting the recA protein to the other genes in the network, bypassing the lexA master regulator. This result can be explained by the one-step time delay inherent in the VAR(1) simulator of the ABC-Net method. More specifically, when DNA damage in the cell is detected by recA, the abundance of lexA decreases very rapidly and the remaining S.O.S. genes turn on almost immediately. However, time-delay models (like the VAR(1) simulator) are only able to identify gene-to-gene interactions that occur with a one-step time lag. The result of this is that in the findings of the ABC-Net method, a strong link appears to occur directly between recA and the remaining genes in the network.

%%%%%%%%%%%%%%%%%%%%%%%%%%%%%%%%%%%%%%%%%%%%%%%%%%%%%%%%%%%%%%%%%%%%%%%%%%%%%%%%%%%%%%%%%%%%%%%%%%%%%%%%%%%%%%
%%%%%%%%%%%%%%%%%%%%%%%%%%%%%%%%%%%%%%%%%%%%%%%%%%%%%%%%%%%%%%%%%%%%%%%%%%%%%%%%%%%%%%%%%%%%%%%%%%%%%%%%%%%%%%

\section{Discussion}
\label{s:discuss}

Reverse engineering the structure of GRN from longitudinal expression data is an intrinsically difficult task, given the complexity of network architecture, the large number of potential gene-to-gene interactions in typical networks, and the small number of replicates and time points available in real data. In this work, we proposed a non-standard extension of the existing ABC-MCMC method \citep{Marjoram2003} to enable inference of GRN. Based in approximate Bayesian computation, the ABC-Net approach enables Bayesian inference for complex, high-dimensional networks for which the likelihood is difficult to calculate. By sampling from the approximate posterior distributions of parameters involved in GRN, this method yields a wealth of information about the structure and inferability of complicated biological systems, particularly with respect to the flexibility and rigidity of network interactions. For the time being, the complexity of real biological systems and the computing time required for the ABC-Net limits its application to small networks. 

As noted by previous authors \citep[e.g.,][]{Sisson2007, Wegmann2009}, there are a number of drawbacks to the ABC-MCMC algorithm. For example, the choice of $\epsilon$ plays an important role in the chain; too large of a value for the threshold $\epsilon$ results in a chain dominated by the prior distribution, while too small of a value leads to extremely low acceptance rates. As such, implementation of the ABC-Net method requires some user tuning. In addition, the number of steps required in the burn-in period and in the chain itself are also dependent on this threshold value. Further work is required to fully examine the components of the ABC-Net method, including more efficient network structure proposal schemes, and techniques to identify optimal data simulators for real data. In particular, a key aspect in this work is the choice of the model used to generate pseudo data; recent advances in using ABC algorithms for parameter inference and as an exploratory tool for model assessment \citep{Ratmann2011} may be useful for this purpose.

In this work, we have suggested the use of somewhat loosely defined ``flexible" and ``rigid" gene-to-gene interactions to better understand the inferability of gene regulatory networks; additional work is required to determine an objective criterion to characterize this behavior. Although our implementation of ABC methods for reverse engineering GRN was a first attempt to demonstrate the flexibility and potential of this procedure, some improvements in performance and efficiency can be expected from the implementation of new simulation techniques, such as population and sequential Monte Carlo \citep{DelMoral2006, DelMoral2009, Robert2010}. Finally, a substantial advantage of the ABC-Net method is its capacity to analyze time-series digital gene expression measurements (e.g., serial analysis of gene expression or RNA sequencing data) through a simple modification of the data simulator (e.g., an autoregressive simulator for Poisson distribution rates). To this end, additional work is required to determine the most appropriate techniques for simulating time-series count data, as well as distance functions best adapted to time-series count data. This goal is particularly important, as the decreasing cost and refinement of next-generation sequencing technology ensure that longitudinal gene expression profiles will likely be studied using RNA sequencing methodology in the near future. 

%%%%%%%%%%%%%%%%%%%%%%%%%%%%%%%%%%%%%%%%%%%%%%%%%%%%%%%%%%%%%%%%%%%%%%%%%%%%%%%%%%%%%%%%%%%%%%%%%%%%%%%%%%%%%%
%%%%%%%%%%%%%%%%%%%%%%%%%%%%%%%%%%%%%%%%%%%%%%%%%%%%%%%%%%%%%%%%%%%%%%%%%%%%%%%%%%%%%%%%%%%%%%%%%%%%%%%%%%%%%%

\section*{Appendix}
Let $\mathbf{y}$ and $\mathbf{y}^\star$ denote observed and simulated time-course expression data, and let $T$ and $P$ denote the number of time points collected and total number of genes, respectively. The Canberra, Euclidean ($L^2$), and Manhattan ($L^1$) distances, respectively, may be defined as 
\begin{align*}
\rho(\mathbf{y}^\star, \mathbf{y}) &= \displaystyle\sum\limits_{t=1}^T \displaystyle\sum\limits_{i=1}^P \frac{\vert y_{it}^\star - y_{it}\vert}{\vert y_{it}^\star + y_{it}\vert} \\
\rho(\mathbf{y}^\star, \mathbf{y}) &= \sqrt{\displaystyle\sum\limits_{t=1}^T \displaystyle\sum\limits_{i=1}^P (y_{it}^\star - y_{it})^2} \\
\rho(\mathbf{y}^\star, \mathbf{y}) &= \displaystyle\sum\limits_{t=1}^T \displaystyle\sum\limits_{i=1}^P \vert y_{it}^\star - y_{it} \vert.
\end{align*}
In addition, we also apply a distance measure proposed by \citet{Lund2009} tailored to multivariate longitudinal data that we refer to as the Multivariate Time-Series (MVT) distance. For the MVT distance, under the null hypothesis that $Y^\star$ and $Y$ have the same network dynamics, we define 
{\allowdisplaybreaks
\begin{align*}
&\hat{\Theta}_{\mathbf{y}} = \frac{1}{T}\sum_{t=1}^{T-1} \mathbf{y}_{t+1}\mathbf{y}_t^{\prime} \\
&\hat{\Theta}_{\mathbf{y}^\star} = \frac{1}{T}\sum_{t=1}^{T-1} \mathbf{y}_{t+1}^\star\mathbf{y}_t^{\star\prime} \\
&\hat{\Theta} = \frac{\hat{\Theta}_{\mathbf{y}}+\hat{\Theta}_{\mathbf{y}^\star}}{2} \\
&\hat{\mathbf{y}}_t^\star = \hat{\Theta}\mathbf{y}_{t-1}^\star \\
&\hat{\mathbf{y}}_t = \hat{\Theta}\mathbf{y}_{t-1} \\
&\hat{\Sigma} = \frac{1}{2T}\sum_{t=1}^T \left \{ (\mathbf{y}_t^\star-\hat{\mathbf{y}}_t^\star)(\mathbf{y}_t^\star-\hat{\mathbf{y}}_t^\star)^\prime + (\mathbf{y}_t-\hat{\mathbf{y}}_t)(\mathbf{y}_t-\hat{\mathbf{y}}_t)^\prime \right \}
\end{align*}}
where $\mathbf{y}_t$ and $\mathbf{y}_t^\star$ are the observed and simulated time-course data, $\hat{\mathbf{y}}_t$ and $\hat{\mathbf{y}}_t^\star$ are the best one-step ahead linear predictors of $\mathbf{y}_t$ and $\mathbf{y}_t^\star$, respectively, and $\hat{\Sigma}$ is an estimate of the common covariance matrix of the errors $\Sigma$. With these terms defined, the MVT distance may be defined as follows:
\begin{align*}
\rho(\mathbf{y}^\star, \mathbf{y}) &= \frac{1}{T}\displaystyle\sum\limits_{t=1}^T \left[(\mathbf{y}_t - \mathbf{y}_t^\star) - (\hat{\mathbf{y}}_t - \hat{\mathbf{y}}_t^\star) \right]^\prime \hat{\Sigma}^{-1} \times\\ 
& \times \left[(\mathbf{y}_t - \mathbf{y}_t^\star) - (\hat{\mathbf{y}}_t - \hat{\mathbf{y}}_t^\star) \right]. \nonumber
\end{align*}

\begin{acknowledgements}
We gratefully acknowledge discussions with Professors Alan Qi, Bruce Craig, and Jayanta Ghosh, as well as helpful comments from two anonymous reviewers that greatly improved this paper. We also thank My Truong and Doug Crabill for their computing expertise, and Paul L. Auer for his careful reading of the manuscript. This research is supported by a grant from the National Science Foundation Plant Genome (DBI 0733857) to RWD.
\end{acknowledgements}

\bibliographystyle{spmpscinat}       % mathematics and physical sciences
\bibliography{ABCNetRef}   % name your BibTeX data base

\end{document}